\documentclass[12pt,preprint]{aastex}
\input colordvi
\newcommand{\ltwid}{\mathrel{\raise.3ex\hbox{$<$\kern-.75em\lower1ex\hbox{$\sim$
}}}}
\newcommand{\gtwid}{\mathrel{\raise.3ex\hbox{$>$\kern-.75em\lower1ex\hbox{$\sim$
}}}}
\newcommand{\bd}{\begin{description}}
\newcommand{\ed}{\end{description}}
\newcommand{\pr}{{\s\prime}}
\newcommand{\be}{{\bf e}}
\newcommand{\s}{\scriptscriptstyle}
\begin{document}
\title{The Proximity of Mercury's Spin to Cassini State 1}
\author{S. J. Peale}
\affil{Department of Physics\\
University of California\\
Santa Barbara, CA 93106\\
peale@io.physics.ucsb.edu}
\begin{abstract}
In determining Mercury's core structure from its rotational
properties, the value of the normalized moment of inertia, $C/MR^2$,
from the location of Cassini 1 is crucial. If Mercury's spin axis
occupies Cassini state 1, its position defines the location of the
state, where the axis is fixed in the frame precessing with the orbit.
Although tidal and core-mantle dissipation drive the spin to the Cassini state
with a time scale $O(10^5)$ years, the spin might still be displaced from
the Cassini state if the variations in the orbital elements induced by
planetary perturbations, which change the position of the Cassini
state, cause the spin to lag behind as it attempts to follow the state.
After being brought to the state by dissipative processes, the spin
axis is expected to follow the Cassini state for orbit variations with 
time scales long compared to the 1000 year precession period of the
spin about the Cassini state because the solid angle swept out by the
spin axis as it precesses is an adiabatic invariant.  Short period
variations in the orbital elements of small amplitude should cause
displacements that are commensurate with the amplitudes of the short
period terms.  The exception would be if there are forcing terms in
the perturbations that are nearly resonant with the 1000 year
precession period. The precision of the radar and eventual spacecraft
measurements of the position of Mercury's spin axis warrants a check on
the likely proximity of the spin axis to the Cassini state. How
confident should we be that the spin axis position defines the Cassini
state sufficiently well for a precise determination of $C/MR^2$? 

By following simultaneously the spin position and the Cassini state
position during long time scale orbital variations over past 3 million
years (Quinn {\it et al.}, 1991) and short time scale variations for
20000 years (JPL Ephemeris DE 408, E. M. Standish, private
communication, 2005), we show that the 
spin axis will remain within one arcsec of the Cassini state after it
is brought there by dissipative torques.  In this process the spin is
located in the orbit frame of reference, which in turn is referenced
to the inertial ecliptic plane of J2000. There are no perturbations
with periods resonant with the precession period that could
cause large separations. We thus expect Mercury's spin to occupy
Cassini state 1 well within the uncertainties for both radar and
spacecraft measurements, with correspondingly tight constraints on
$C/MR^2$ and the extent of Mercury's molten core.  Two unlikely
caveats for this conclusion are 1. an excitation of a free spin precession
by an unknown mechanism or 2. a displacement by a dissipative core mantle
interaction that exceeds the measurement uncertainties. 
\end{abstract}
\section{Introduction}
Radar observations have begun the process of determining the
extent of Mercury's molten core (Margot et al. 2003), and the
experiment  will be completed when the MESSENGER spacecraft orbits Mercury
in 2011 (Solomon et al. 2001). The BepiColombo spacecraft will
complement and augment the observations of MESSENGER, but will orbit
Mercury sometime after 2011 (Anselmi and Scoon, 2001).   The experiment
is based on the product of three factors (Peale, 1976; 1981; 1988;
2005; Peale et al. 2002). 
\begin{equation}
\left(\frac{C_m}{B-A}\right)\left(\frac{MR^2}{C}\right)\left(\frac{B-A}
{MR^2}\right)=\frac{C_m}{C}\leq 1, \label{eq:cmoverc}
\end{equation}
where $A<B<C$ are the principal moments of inertia of Mercury with
$C_m$ being the polar moment of inertia of the mantle alone, and $M$ and
$R$ are Mercury's mass and radius respectively.  The first factor is
determined by the amplitude of the physical libration,
$\phi_0=[3(B-A)/2C_m](1-11e^2+\cdots),$ where $e$ is the orbital
eccentricity and $C_m$ appears in the denominator because the liquid
core is not expected to follow the mantle during the short period
librations (Peale, et al. 2002). The second factor follows from the analysis
of generalized Cassini's laws for Mercury (Colombo, 1966; Peale, 1969;
Beletskii, 1972) (see Section \ref{sec:cassinistate}).
\begin{equation}
\frac{C}{MR^2}=\frac{nJ_2}{w_{\s L}}\frac{f(e)}
{(\sin{I^\pr})/i_c+\cos{I^\pr}},\label{eq:cmr2}
\end{equation}
where $f(e)=G_{210}(e)+2C_{22}G_{201}(e)/J_2$, $n=\sqrt{GM_\sun/a^3}$ is
the orbital mean motion, ($G$ is
the gravitational constant, $M_\sun$ is the  
solar mass, $a$ is the semimajor axis of the orbit), the
functions $G_{210}(e)=(1-e^2)^{-3/2}$ and $G_{201}(e)=7e/2-
123e^3/16+...$ are defined by Kaula (1966), $J_2$ and
$C_{22}$ are the second degree coefficients in the  
expansion of Mercury's gravitational field, $I^\pr$ is the inclination of
the orbit plane to the Laplace plane, the plane on which Mercury's
orbit precesses with nearly constant inclination and at a nearly
constant rate whose magnitude is $w_{\s L}$, and $i_c$ is the obliquity of
the Cassini state. The 
orientation of the Laplace plane and the value of $w_{\s L}$ are determined
by averaging the orbital element variations over a suitable time
interval as discussed below. The last factor is found from
$C_{22}=(B-A)/(4MR^2)$.   

\begin{figure}[h]
\epsscale{0.7}
\plotone{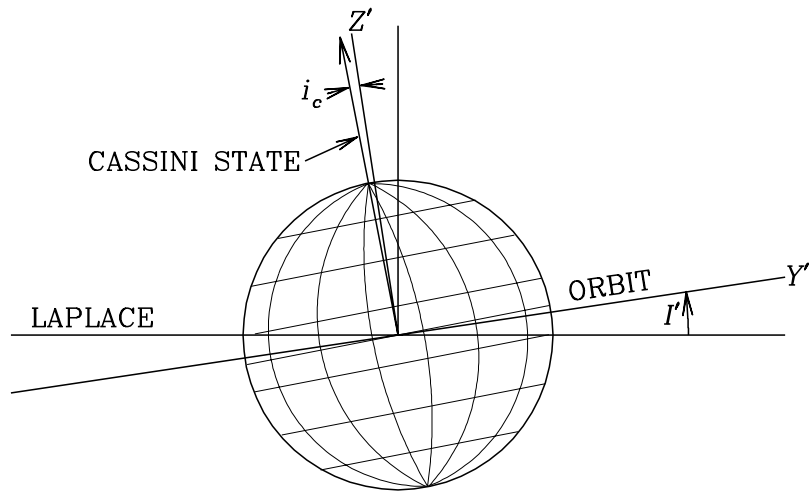}
\caption{Geometry of Cassini state 1. The ascending node of the
equator on the orbit plane and the ascending node of the orbit on the
Laplace plane remain coincident as they precess around the normal to
the Laplace plane at the current rate corresponding to the orbit precession
period of about 300,000 years. \label{fig:cassinistate}}
\end{figure}

Fig. \ref{fig:cassinistate} shows the geometry
of Cassini state 1 for Mercury, where Mercury's obliquity $i=i_c$ with
the ascending nodes of the equator plane on the orbit plane and of the
orbit plane on the Laplace plane remaining coincident as the spin and
orbit normal precess around the normal to the Laplace plane. 
The Laplace plane is determined mainly by Venus, Earth
and Jupiter as is the precession of Mercury's orbit. Because of the
orbit precession, the spin precesses around the Cassini state and
tends toward that state from dissipative effects rather than toward
the orbit normal (Peale, 1974: Ward, 1975).

The determination of $\phi_0,\,J_2,\,C_{22},\,{\rm and}\,i_c$ thus yields 
$C_m/C$.  Empirical constraints on $C/MR^2$ must wait for MESSENGER
determinations while in orbit.  Crucial to the determination of moment
of inertia is the assumption that the spin axis occupies Cassini state 1
(Eq. (\ref{eq:cmr2})). Only if we are confident that the measured
obliquity $i=i_c$, can we constrain the value of $C/MR^2$ with sufficient
precision to determine the extent of Mercury's liquid
core. Knowing the internal structure of Mercury is important for
inferring the thermal history, overall chemical composition and the
constraints Mercury can place on the details of origin and evolution
of the terrestrial planets (Harder and Schubert, 2001; Solomon et
al. 2001). 

Tidal friction will drive Mercury's spin to Cassini state 1 from
virtually any initial obliquity (Peale, 1974; Ward, 1975), and the
addition of dissipation between a liquid core and solid mantle will
hasten that evolution, provided the initial obliquity $i<90^\circ$,
with a time scale from both tides and 
core-mantle dissipation together of order $10^5$ years (Peale,
2005).  As Mercury's spin axis approaches the Cassini state, it will
precess around that state with a period near 500 or 1000 years 
depending on whether or not the liquid core is dragged along with the
spin precession.  A finite amplitude free precession would
frustrate the determination of $i_c$, but the short time scale
for its decay makes a remnant free spin precession unlikely, unless there
is a recent, unspecified excitation mechanism (Peale, 2005).   

Another possibility for a displacement of the spin from the
Cassini state is if the spin is unable to follow the Cassini state
sufficiently closely as the position of the state changes due to the
variations in the orbital elements and Laplace plane orientation
from planetary perturbations and changing solar system geometry.
We can expect the spin to remain reasonably close to the Cassini state
after being brought there by dissipative processes because the
following action  integral is an adiabatic invariant, where the
integral is nearly constant if the period of
the angle variable is short compared to all other relevant time scales
(Goldreich and Toomre, 1969; Peale, 1974).  
\begin{equation}
\oint{p^\pr dq^\pr}=-\alpha\oint(1-\cos{\theta^\pr})d\phi^\pr\approx
{\rm const},\label{eq:action}
\end{equation}
where the action variable $p^\pr=-\alpha(1-\cos\theta^\pr)$ with
$\alpha$ being the spin angular momentum, and $q^\pr=\phi^\pr$ is the
angle variable, with $\theta^\pr$ and $\phi^\pr$ being the ordinary
spherical polar coordinates of the spin vector in the system with the
$Z$ axis aligned with the Cassini state. That $p^\pr$ and $q^\pr$ are
conjugate variables is 
verified by $dp^\pr/dt=-\partial H/\partial q^\pr$ and
$dq^\pr/dt=\partial H/\partial p^\pr$, with $H$ being the Hamiltonian
for the Mercury's rotational dynamics written in the frame precessing
with the orbit (Peale, 1974).  The action integral is seen to be
$-\alpha$ times the solid angle contained by the spin vector as it
precesses around the Cassini state,  and it is approximately conserved
if the spin precession rate $\dot\phi^\pr$ is fast relative to the
significant changes in the parameters determining the position of the
state. The angle variable describes the precession of the spin about
the Cassini state.  We shall see below that the large amplitude
variations in the orbital parameters relative to the ecliptic plane
have periods exceeding $5\times 10^4$ years, which is sufficiently
long compared to the 1000 year precession period that one expects the
adiabatic invariant to keep the spin close to the current position of 
the Cassini state (once it is there) as the latter's position changes
slowly on these 
time scales. The adiabatic invariant is not conserved on the time
scale of short period variations, but these variations are of small
amplitude, and we shall see that the spin follows the Cassini state
defined by the orbital elements averaged in a 2000 year window over
the 20,000 year JPL Ephemeris DE 408.
The precision of the radar determinations of Mercury's spin
properties, and that anticipated for the MESSENGER and BepiColombo
missions warrants a check on just how well the spin axis follows the
Cassini state for all variations of the parameters defining the state.

Our purpose here is to develop a formalism to test how close 
to the Cassini state the spin axis remains for both slow and fast  
variations in the orbital parameters.  We shall consider the
variations of  $e,\,I,\,\Omega$, and their derivatives due to the
planetary perturbations, where $I$ and $\Omega$ are the are the orbit
inclination and longitude of the ascending node of the orbit on the
ecliptic plane respectively. Dissipative processes will  
relentlessly drive the spin to the Cassini state, so the test will determine
how closely Mercury's spin axis follows the Cassini state position as
the position of the state changes due to the orbital element variations.
To this end we develop the equations of motion of
Mercury's spin vector relative to the orbit frame of reference,
averaged over an orbit period, in Section \ref{sec:eqns}, and
find the position of the spin and the position of the Cassini state as
a function of time as variations in the orbital parameters
ensue.  Finding the Cassini state position will involve defining a set
of coplanar vectors that includes the normal to the
Laplace plane. The inclination of the orbit relative to the Laplace
plane and the rate of precession of the ascending node of the orbit on the
Laplace plane do not have to be known with high precision to define
the position of the Cassini state accurately (Yseboodt and Margot,
2005).  For the slow variations in 
$e,\,I,\,\Omega,\,dI/dt$ and $d\Omega/dt$, we shall vary the orbital elements
and their time derivatives according to simulations by T. Quinn (Quinn et 
al. 1991) over the past $3\times 10^6$
years.\footnote{ftp://ftp.astro.washington.edu/pub/hpcc/QTD}. For the 
short period variations in the orbital elements we follow the spin and
Cassini state positions as $e,\,I,\,\Omega,\, dI/dt$ and $d\Omega/dt$ 
vary according to the 20,000 year JPL Ephemeris DE 408 provided by
Myles Standish, but now the Cassini state position will be determined
by elements and rates averaged over the 2000 year window mentioned above.
Periodic variations of the same variables with amplitudes
and periods representative of the true variations are also applied to
identify the sources of the fluctuations in the spin-Cassini state
separation that are observed.

In Section \ref{sec:cassinistate} we determine the time varying
position of the Cassini state, where variations in
$dI/dt$ and $d\Omega/dt$ change the Laplace plane orientation and the
variations in $I,\, e$ and $w$ change $i_c$. This 
determination allows us to compare the position of the spin vector
with the Cassini state position at any time.  All 
dissipative forces are ignored, and principal axis rotation is
assumed. The neglect of dissipation means we couple the liquid core
firmly to the mantle, which is equivalent to Mercury's having a solid
core.  Consequences of the relaxation of this latter assumption will
be pursued in a later paper.  

In Section \ref{sec:quinn}, the spin vector, placed initially in the
Cassini state,  
is shown to remain within approximately $1^{\pr\pr}$ of the Cassini
state position as this position is continuously redefined for the
continuously varying Laplace plane and orbital elements as given over
the past $3\times 10^6$ years by the Quinn simulation. Equivalently,
the spin vector would maintain any small initial separation from the 
Cassini state to within $1^{\pr\pr}$. The effect of higher frequency
terms on the spin-Cassini state separation is found in Section
\ref{sec:standish}. Large separations of the spin from the
Cassini state are forced for periodic variations in the inclination 
with periods near the spin precession period.  However, any terms with
periods near resonance with the spin precession period 
are not evident in the real variations of $e,\,I,\,\Omega,\, dI/dt$ and
$d\Omega/dt$ . This is demonstrated by the spin and Cassini state
positions (from locally averaged values of elements and rates)
remaining within $1^{\pr\pr}$ for variations of these parameters given
by JPL Ephemeris DE 408. A summary and discussion follows in Section
\ref{sec:discussion}, where we end with a conjecture on the possible
effect of the liquid core on the position of the spin axis.      

\section{Equations of variation \label{sec:eqns}}

\begin{figure}[h]
\epsscale{.7}
\plotone{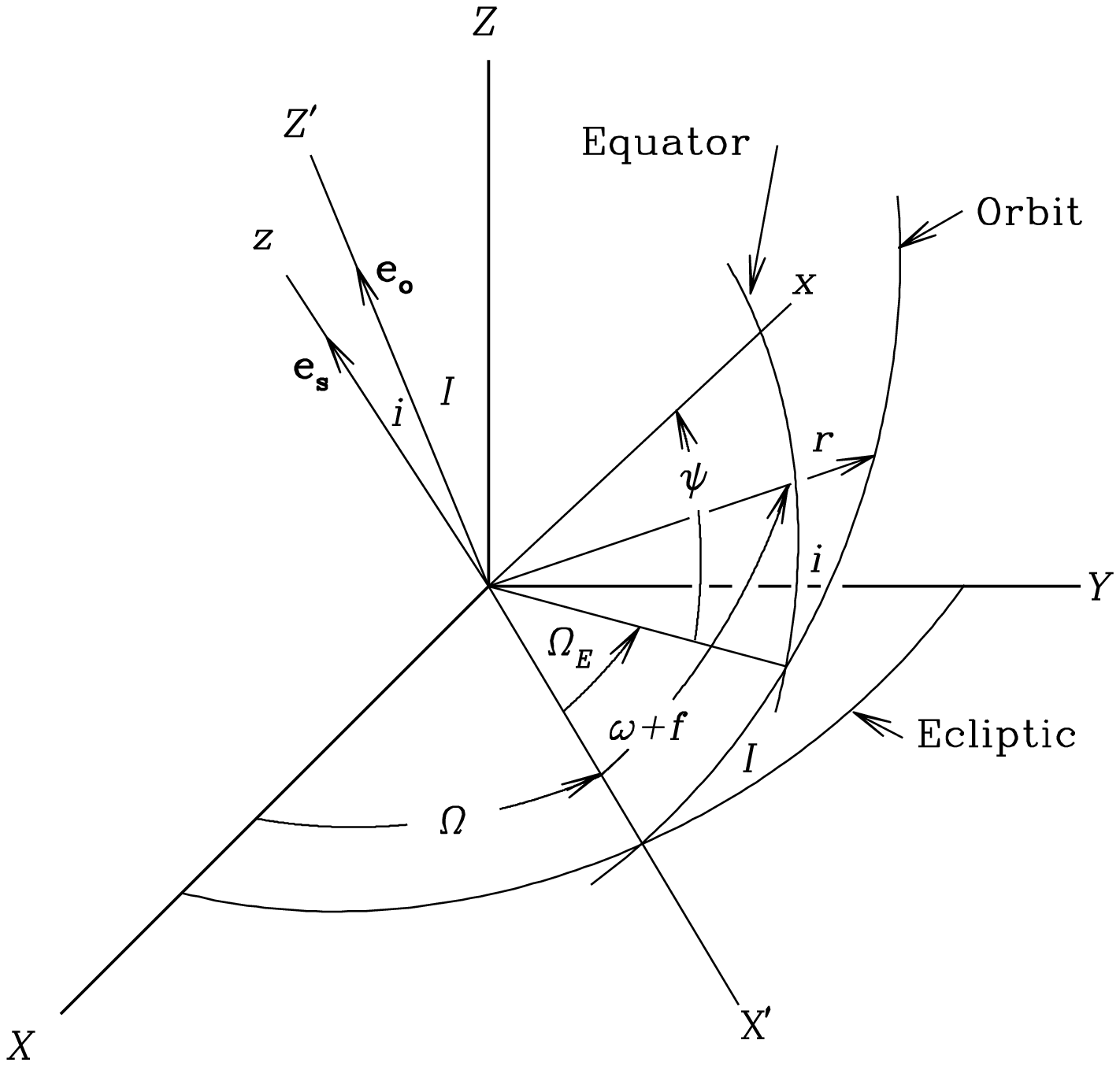}
\caption{Coordinate systems used in determining time variation of
Mercury's spin vector, where $\be_s$ and $\be_o$ are unit vectors
along the spin axis and orbit normal respectively. The $XY$ inertial
plane is the ecliptic plane of J2000. \label{fig:lplane}}
\end{figure}

Fig. \ref{fig:lplane} defines the coordinate systems centered on
Mercury to be used along with some of the variables. The $XYZ$ system
is the inertial ecliptic plane of J2000 with the $Z$ axis
perpendicular to the plane. The $X^\pr Y^\pr Z^\pr$ system has the
$Z^\pr$ axis perpendicular to the orbit  plane and the $X^\pr$ axis
along the ascending node of the orbit plane on the XY plane. The $xyz$
system is the principal axis system fixed in the body. The orbit plane
is inclined to the $XY$  plane by angle $I$, $\Omega$ is the
longitude of the ascending node of the orbit plane on the $XY$
plane measured from the inertial $X$ axis along the direction to the
vernal equinox, $\be_o=\be_{\s Z^\pr}$ is the orbit normal,
$\be_s=\be_z$ is a unit vector along the spin axis, $\Omega_{\s E}$ is the 
longitude of the ascending node of Mercury's equator on the orbit
plane relative to the $X^\pr$ axis, $i$ is inclination of the equator
plane relative to the orbit plane (obliquity), $\psi$ is the angle between
the ascending node of the equator and the $x$ principal axis, $\bf r$
points toward the Sun in the orbit plane and $\omega$ and $f$ are the
argument of perihelion and true anomaly locating the Sun in the orbit
plane relative to the $X^\pr$ axis. 

In Peale (2005) it is shown that although the actual spin precession
trajectory is slightly elliptical and the rate of precession is not
quite uniform, a good approximation for the time variation of the spin
vector is given by 
\begin{equation}
\frac{d\be_s}{dt}=K_1\cos{i}\,(\be_s\times \be_o)=-K_1\cos{i}
\sin{i}(\cos{\Omega_{\s E}}\,\be_{\s X^\pr}+ \sin{\Omega_{\s
E}}\,\be_{\s Y^\pr})
\label{eq:desdt} 
\end{equation}
where 
\begin{equation}
K_1=\frac{n^2MR^2}{C\dot\psi}\left[\frac{3}{2}J_2G_{210}(e)
+3C_{22}G_{201}(e)\right]=Kf(e), 
\label{eq:K1}
\end{equation}
with $K=nMR^2J_2/C$ and
$f(e)=G_{210}(e)+2C_{22}G_{201}(e)/J_2$, (used in Eq. (\ref{eq:cmr2}))
and with $\dot\psi=3n/2$
being assumed. 
Eq. (\ref{eq:desdt}) follows from the average of the equations of motion
over the orbit period as mentioned above with $\dot\psi=1.5n$, and
neglect of terms with coefficients that are small compared with
Eq. (\ref{eq:K1}). We can consider only the variation of the unit
vector $\be_s$ because $\dot\psi\equiv 1.5n$ from the spin orbit
resonance if we neglect the small physical librations.
Eq. (\ref{eq:desdt}) shows the expected regression 
of the spin vector about the orbit normal. It is easiest to determine
the equations of motion of the spin in the orbit frame of reference
starting from the relation    
\begin{equation}
\frac{d\be_s}{dt}_{inertial}=\frac{d\be_s}{dt}_{orbit}+\vec w\times
\be_s =K_1\cos{i}\,(\be_s\times \be_o),\label{eq:desdt2}
\end{equation}
where $\be_s=\sin{i}\sin{\Omega_{\s E}}\,\be_{\s X^\pr}-\sin{i}\cos{\Omega_{\s
E}}\,\be_{\s Y^\pr}+\cos{i}\,\be_{\s Z^\pr}$  and $\be_o=\be_{\s Z^\pr}$
are defined in Fig. \ref{fig:lplane}. The $\be_i$ are unit vectors
along the respective axes. 

The angular velocity $\vec w$ is nominally $\vec w_{\s L}$, the
angular velocity of the orbit normal about the Laplace plane
normal, but we can make another choice that simplifies the equations. 
The precession of the orbit normal about the Laplace plane normal 
leads to a velocity of the orbit normal $\be_o$ of ${\vec v}=\vec w_{\s
L}\times \be_o$, where $\vec w_{\s L}$ is the angular velocity of the orbit
normal parallel to the Laplace plane normal.  Differentiation of the
components of $\be_o$ in the ecliptic frame yields components of
${\vec v}$ that can be set 
equal to the like components of the cross product.   However,
these equations for the components of $\vec w_{\s L}$ are not linearly
independent, so one can only solve for two of the components in terms of
the other, which serves as a free parameter. A convenient choice of
the free parameter is the $Z$ component of $\vec w$, where we have
abandoned the subscript, since another constraint is necessary to
determine the Laplace plane normal.
\begin{equation}
 \vec w=\left[\frac{dI}{dt}\cos{\Omega}+(w_{\s
 Z}-\frac{d\Omega}{dt})\tan{I}\sin{\Omega}\right]\be_{\s X}
 +\left[\frac{dI}{dt}\sin{\Omega}+(-w_{\s
 Z}+\frac{d\Omega}{dt})\tan{I}\cos{\Omega}\right]\be_{\s Y}
 +w_{\s Z}\be_{\s Z}. \label{eq:vecw1}
\end{equation}

All of the solutions for $\vec w$ lie in the plane determined by the
orbit normal and the Laplace plane normal, where the latter is one of
the set. The coplanarity is understood since all the $\vec w\times
\be_o$ must produce the same ${\vec v}$.  If $I^\pr$ is the angle
between $\be_o$ and $\vec w$, $v=w\,\sin{I^\pr}$ is a constant, and
the magnitude $w$ must decrease as $I^\pr$ increases.
Eq. (\ref{eq:vecw1}) is Eq. (13) of Yseboodt and Margot (2005), where
they determine $\vec w_{\s L}$ at a given epoch by adding the
numerically determined constraint that the variation in the
inclination $I^\pr$ is minimized in a 2000 year window centered on the
epoch for data obtained from the 20,000 year JPL Ephemeris DE 408. At
the epoch J2000, the ecliptic latitude and longitude of $\vec w_{\s
L}$ are $86.725^\circ$ and $66.6^\circ$ respectively corresponding to
$w_{\s Z}=-1.91\times 10^{-5}$ radians/year, $I^\pr=8.6^\circ$ and
an instantaneous precession period of 328,000 years. There is some
uncertainty in these values from the statistical nature of the
minimization process, but we shall see below that the uncertainty in
the Laplace plane normal does not compromise the determination of the
Cassini state.  

\begin{figure}[ht]
\epsscale{.7}
\plotone{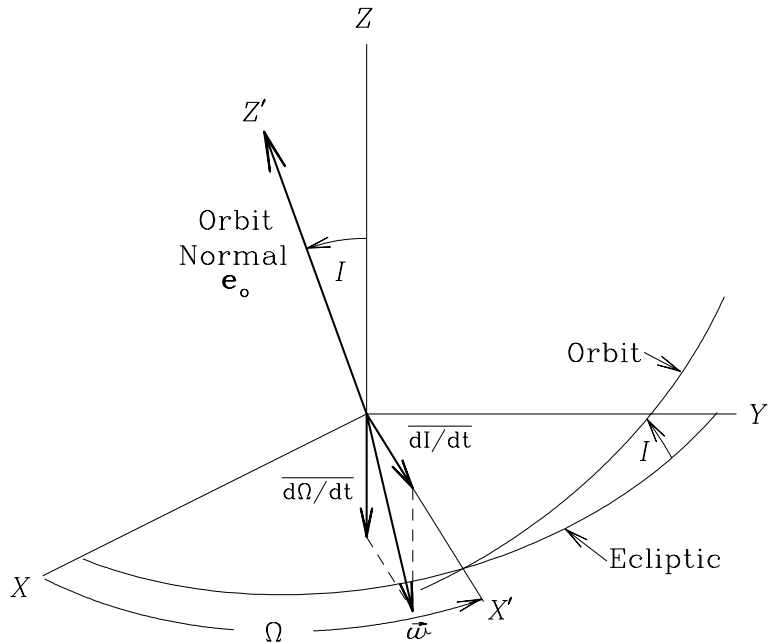}
\caption{The vector sum of $\overline{d\Omega/dt}$ and
$\overline{dI/dt}$ yields an effective precessional angular velocity
$\vec w$ of the orbit normal, which is the simplest choice from
Eq. (\ref{eq:vecw1}). 
\label{fig:dIdtprecession}} 
\end{figure}

The simplest form for $\vec w$ in Eq. (\ref{eq:vecw1}) is for $w_{\s
Z}=d\Omega/dt$.  Since any choice of $w_{\s Z}$ in
Eq. (\ref{eq:vecw1}) yields the correct
instantaneous motion of $\be_o$, we can use this simplest form, 
\begin{equation}
\vec w=\frac{dI}{dt}\be_{\s X^\pr}+\sin{I}\frac{d\Omega}
{dt}\be_{\s Y^\pr}+\cos{I}\frac{d\Omega}{dt}\be_{\s Z^\pr},
\label{eq:vecomega}
\end{equation}
in Eq. (\ref{eq:desdt2}) to determine the equations of motion of the
spin in the orbit frame of reference. This choice of $w$, expressed
here in the orbit frame, is just the vector sum of $\overline{dI/dt}$
and $\overline{d\Omega/dt}$ (Fig. \ref{fig:dIdtprecession}). 
Equating like components in Eq. (\ref{eq:desdt2}) yields three linearly
dependent equations that can be solved uniquely for $di/dt$ and
$d\Omega_{\s E}/dt$. There results
\begin{eqnarray}
\frac{di}{dt}&=&-\sin{I}\sin{\Omega_{\s E}}\frac{d\Omega}
{dt}-\cos{\Omega_{\s E}}\frac{dI}{dt}\nonumber\\
\frac{d\Omega_{\s E}}{dt}&=&-K_1\cos{i}+
\frac{\cos{i}\sin{\Omega_{\s E}}}{\sin{i}}\frac{dI}{dt} \nonumber\\
&&-\frac{\cos{i}\cos{\Omega_{\s E}}\sin{I}+\sin{i}\cos{I}}{\sin{i}}
\frac{d\Omega}{dt}, \label{eq:eqmotion}
\end{eqnarray}
where $dI/dt$ and $d\Omega/dt$ are assumed known. 

As expected, $di/dt=0$ and $d\Omega_{\s
E}/dt=-K_1\cos{i}=$ constant if $dI/dt$ and
$d\Omega_{\s L}/dt$ are both zero. If $dI/dt=0$ but $d\Omega/dt$ is a
negative constant, the equations describe the motion of 
the spin precession about a Cassini state in a uniformly precessing
orbit. The Cassini state is displaced from the orbit normal
in the plane defined by the orbit normal $\be_o$ and the 
normal $\be_{\s L}=-\vec w_{\s L}/w_{\s L}$
(Fig. \ref{fig:cassinistate}) (Colombo 1966, Peale, 1969). This is the
same plane containing all of the vectors defined by
Eq. (\ref{eq:vecw1}).    

To eliminate the $\sin{i}$ singularity in Eqs. (\ref{eq:eqmotion})
we introduce the variables $p=\sin{i}\sin{\Omega_{\s E}}$ and
$q=\sin{i}\cos{\Omega_{\s E}}$, with the result
\begin{eqnarray}
\frac{dp}{dt}&=&-Kf(e) q\sqrt{1-p^2-q^2}-
(\sin{I}\sqrt{1-p^2-q^2}+q\cos{I})\frac{d\Omega}{dt},\nonumber\\ 
\frac{dq}{dt}&=&Kf(e) p\sqrt{1-p^2-q^2}
-\sqrt{1-p^2-q^2}\frac{dI}{dt}+p\cos{I}\frac{d\Omega}{dt}.
\label{eq:dpqdt}   
\end{eqnarray}
The numerical solution of Eqs. (\ref{eq:dpqdt}) yields the position of
the spin vector of Mercury in the
$X^{\s\prime}Y^{\s\prime}Z^{\s\prime}$ orbit system when all of $e,\, I$ 
and $\Omega$ are varying. 

\section{Cassini state position \label{sec:cassinistate}}
We wish to start the spin axis either in the initial Cassini state or
close to it and determine how close it remains to the changing
position of the state as the spin varies according to
Eqs. (\ref{eq:dpqdt}). 
From Eq. (12) of Peale (1974), the position of the Cassini state 1
is defined by 
\begin{eqnarray}
\sqrt{1-e_{c\s Z^\pr}^2}\cos{I^\pr}+e_{c\s Z^\pr}\sin{I^\pr}-
2R^\pr e_{c\s Z^\pr}\sqrt{1-e_{c\s Z^\pr}^2}-2S(1+e_{c\s Z^\pr})
\sqrt{1-e_{c\s Z^\pr}^2}&=&0\nonumber\\
e_{c\s Z^\pr}&=&0, \label{eq:casstate}  
\end{eqnarray}
where
\begin{eqnarray}
R^\pr&=&\frac{3}{4}\frac{MR^2}{C}\frac{n^2}{\dot\psi w_{\s L}}
J_2G_{210}(e),\nonumber\\ 
S&=&\frac{3}{4}\frac{MR^2}{C}\frac{n^2}{\dot\psi w_{\s L}}
C_{22}G_{201}(e), \label{eq:RS}
\end{eqnarray}
with $e_{c\s X^{\s\prime},Y^{\s\prime},Z^{\s\prime}}$ being the
components in the 
orbit system ($X^{\s\prime}Y^{\s\prime}Z^{\s\prime}$) of $\be_c$,
the unit vector along the Cassini state. For Mercury, Cassini
state 1 is very close to the orbit normal ({\it e.g.}, Peale, 1969,
1974), which geometry is shown in Fig. \ref{fig:cassinistate}.  We can
then set $\sqrt{1-e_{c\s Z^\pr}^2}=|\sin{i_c}|\approx i_c$, $e_{c\s
Z^\pr}\approx 1$ to first order in $i_c$ and write
\begin{equation}
i_c=\frac{w\sin{I^\pr}}{w(2R^\pr+4S-\cos{I^\pr})},
\label{eq:ic} 
\end{equation}
where we have replaced $w_{\s L}$ by $w=\sqrt{(dI/dt)^2+
(d\Omega/dt)^2}$, the magnitude of the vector defined in
Eq. (\ref{eq:vecomega}) and used in the equations of motion of the
spin. The inclination $I^\pr$ in Eq. (\ref{eq:ic}) defined below is
distinct for each choice of $w$. But since we have shown above that
$w\sin{I^\pr}$ has the same value 
for all the values of $w$ compatible with Eq. (\ref{eq:vecw1}), the
only change remaining after this substitution is the change in the
term $w\cos{I^\pr}$ in the denominator.  We shall see below
that $2R^\pr+4S\gtwid 250\gg \cos{I^\pr}$, so there is less than
$\sim 0.1\%$ change in $i_c,\; (\approx 0.05^{\pr\pr}$ at the current epoch)
effected by this substitution. Like the equations of motion for the
spin (Eqs. (\ref{eq:dpqdt})), use of this $w$ in place of $w_{\s L}$
greatly simplifies the numerical determination of the Cassini state
position. 

The inclination of the orbit $I^\pr$ to use in Eq. (\ref{eq:ic}) is
defined by 
\begin{equation}
\frac{-(\be_o\cdot {\vec w})}{w}=\frac{\cos{I}}{\sqrt{1+(dI/dt)^2/
(d\Omega/dt)^2}}=\cos{I^\pr}, \label{eq:cosipr}
\end{equation}
where the minus sign on the left hand side defines $I^\pr$ as a small
positive angle when $d\Omega/dt<0$. Eq. (\ref{eq:ic}) thus defines the
obliquity of the Cassini state in the orbit frame of reference for the
values of $e,\,I,\, dI/dt$ and $d\Omega/dt$ at a particular time. As
noted above, the Cassini state is in the plane defined by
$\be_o=\be_{\s Z^\pr}$ and $\vec w$, which plane also contains $\vec
w_{\s L}$. 

Using Eq. (\ref{eq:vecomega}), we can write (Fig. \ref{fig:lplane}) 
\begin{equation}
-{\vec w}\times{\bf e}_o=-\frac{d\Omega}{dt}\sin{I}\,
{\bf e}_{\s X^{\s\prime}} +
\frac{dI}{dt}\,{\bf e}_{\s Y^{\s\prime}} \label{eq:wcrosseo}
\end{equation}
to define the components of a vector in the orbit ($X^{\s\prime}Y^{\s\prime}
Z^{\s\prime}$) system that is perpendicular to the plane containing
the instantaneous Cassini state 1 and $\vec w$ appropriate to the
instantaneous position of the latter vector. We already have the
obliquity of the Cassini state from Eq. (\ref{eq:ic}), so a unit
vector along the instantaneous Cassini state in the
$X^{\s\prime}Y^{\s\prime} Z^{\s\prime}$ system is
\begin{equation}
{\bf e}_c=\sin{i_c}\sin{\Omega_c}\,{\bf e}_{\s
X^{\s\prime}}-\sin{i_c}\cos{\Omega_c}\,{\bf e}_{\s
Y^{\s\prime}}+\cos{i_c}\,{\bf e}_{\s Z^{\s\prime}},\label{eq:ec}
\end{equation}
where
\begin{eqnarray}
\cos{\Omega_c}&=&\frac{-(d\Omega/dt)\sin{I}}{\sqrt{(d\Omega/dt)^2 
\sin^2{I}+(dI/dt)^2}}, \nonumber\\  
\sin{\Omega_c}&=&\frac{dI/dt}{\sqrt{(d\Omega/dt)^2\sin^2{I}+(dI/dt)^2}},
\label{eq:Omegac}
\end{eqnarray} 
with $\Omega_c$ being the angle between the $X^{\s\prime}$ axis and the
vector in the $X^{\s\prime}Y^{\s\prime}$ plane defined by
Eq. (\ref{eq:wcrosseo}).

\section{Results for slow variations \label{sec:quinn}}
\begin{figure}[ht]
\epsscale{.7}
\plotone{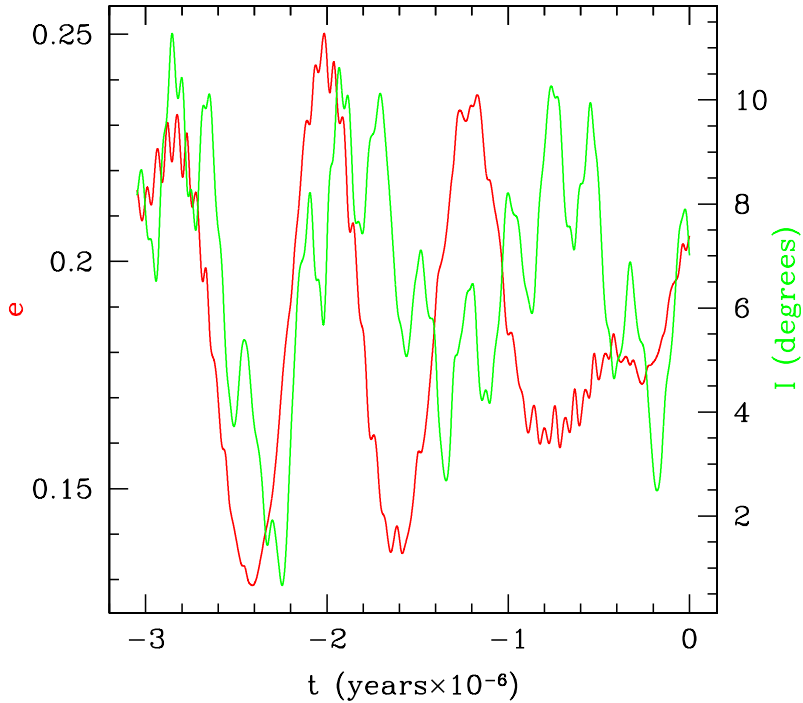}
\caption{Variation of Mercury's eccentricity and orbital inclination to the
ecliptic of J2000 from T. Quinn 
(ftp://ftp.astro.washington.edu/pub/hpcc/QTD).   
\label{fig:quinndata}}
\end{figure}

We are now in a position to check just how well the adiabatic
invariant is satisfied, or more directly, how well the spin follows the
Cassini state.   
In Fig. \ref{fig:quinndata} the variations of Mercury's
eccentricity and inclination to the ecliptic of J2000 are shown
for the last $3\times 10^{6}$ years from the data kindly provided 
by T. Quinn (Quinn, et al. 1991, See footnote $^1$.). These data have
been filtered to exclude any terms with periods less than 2000
years. The justification for this filtering is that such terms usually
are small amplitude oscillations that average to yield negligible
contributions to the element variations.   Note that keeping only 
contributions from terms with periods longer than 2000 years excludes  
any terms close to the spin precession period of between 1000 and 1100
years. The consequences of near resonant terms are discussed in Section
\ref{sec:standish}. 
For the integrations, we divide Eqs. (\ref{eq:dpqdt}) by $K$, defined
after Eq. (\ref{eq:K1}),  such that $Kt$ is a dimensionless
time. The period $2\pi/K\approx 1365$ years would be close to the
spin precession period for small $e$ if $C_{22}=0$. The dimensionless
equations that are integrated are then Eqs. (\ref{eq:dpqdt}) with
$K$ removed from the coefficient of the first term on the rhs  of the
equations, and $Kt\rightarrow t$.  

We note that the spin vector
\begin{eqnarray}
\be_s&=&\sin{i}\sin{\Omega_{\s E}}\,\be_{\s X^\pr}- \sin{i}\cos{\Omega_{\s
E}}\,\be_{\s Y^\pr}+\cos{i}\,\be_{\s Z^\pr},\nonumber\\
&&=p\,\be_{\s X^\pr}-q\,\be_{\s Y^\pr}+\sqrt{1-p^2-q^2}\be_{\s Z^\pr},
\label{eq:es}
\end{eqnarray}
and that Eq. (\ref{eq:ec}) can be written similarly with $p$ and $q$
replaced by $p_c=\sin{i_c}\sin{\Omega_c}$ and
$q_c=\sin{i_c}\cos{\Omega_c}$ with $i_c$ and $\Omega_c$ being defined
in Eqs. (\ref{eq:ic}) and (\ref{eq:Omegac}) respectively. We wish to
determine the angle $\delta$ between $\be_s$ and $\be_c$ as a function
of time, where
\begin{equation}
\cos{\delta}=\be_s\cdot\be_c\approx 1-\frac{\delta^2}{2},
\label{eq:cosdel} 
\end{equation}
so that
\begin{equation}
\delta\approx\sqrt{p^2+q^2+p_c^2+q_c^2-2pp_c-2qq_c} \label{eq:delta}
\end{equation}
gives a more accurate value for $\delta$ than numerically calculating
$\cos^{-1}{\delta}$ when $\delta$ is very small. A cubic spline (Press
et al. 1986) through the Quinn data points is used to determine
$e(t)$, $I(t)$, $\Omega$, $dI/dt$ and $d\Omega/dt$ in the
calculation. The fact 
that contributions to the Quinn data with periods less than 2000 years
have been eliminated, means the values of the variables can be used
directly in calculating the position of the Cassini state without
additional averaging (See Section \ref{sec:standish}).  With the
exception of the starting and ending values used to evaluate the
second derivatives in the spline fit, the values of $dI/dt$ and
$d\Omega/dt$ are determined as defined in the spline algorithm.
\begin{figure}[h]
\epsscale{.6} 
\plotone{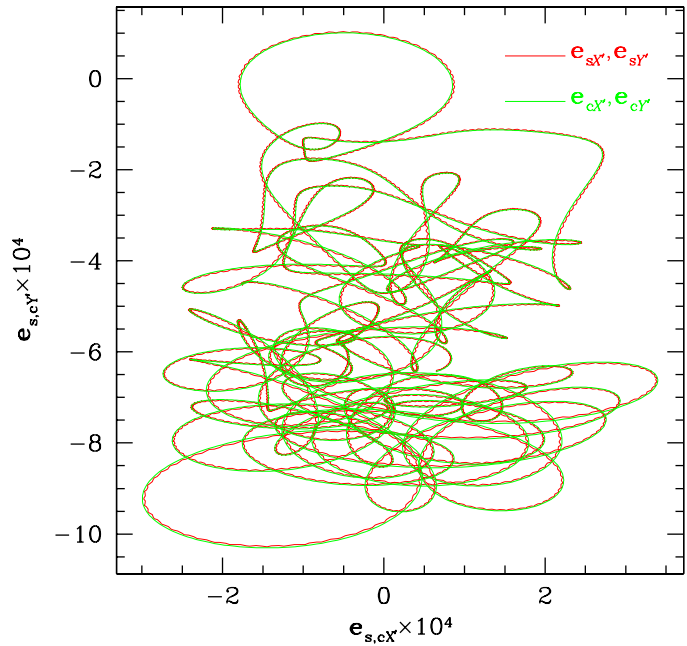}
\caption{Trajectories of the projections of the unit spin vector
$\be_s$ and the unit vector in the Cassini state direction $\be_c$ on
the orbit $X^\pr Y^\pr$ plane for variations in
$e,\,I,\,\Omega,\,dI/dt$ and $d\Omega/dt$  according to
$3\times 10^6$ year simulations by T. Quinn. \label{fig:qp_qcpc3}}
\end{figure}

\begin{figure}[ht]
\epsscale{.6}
\plotone{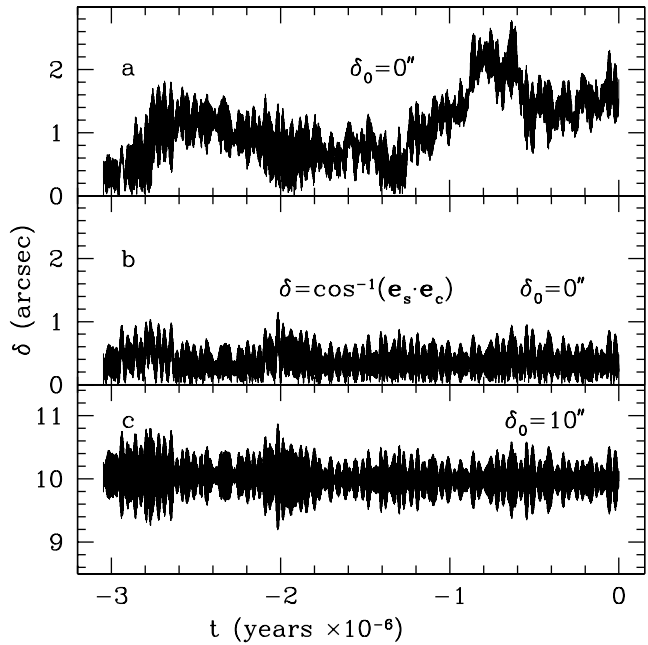}
\caption{Separation $\delta$ of the spin vector from 
 Cassini state 1 for variations in $e,\,I,\,dI/dt$ and $d\Omega/dt$
from a simulation by T. Quinn. The spin is initially at the Cassini
state in panels $a$ and $b$ and separated initially by $10^{\pr\pr}$
in panel $c$. The separation $\delta(t)$ in $a$ is constructed from a spline
fit through all 6185 data points in the Quinn simulation, whereas that
in $b$ results from a spline fit through every other point or 3093
points total. The reason for the differences in $a$ and $b$ is
discussed in the text. Panel $c$ shows the constancy of $\delta$
to within $\pm 1^{\pr\pr}$ for an initial value of $10^{\pr\pr}$.
\label{fig:quinnadiabat4}}
\end{figure}

We show in Fig. \ref{fig:qp_qcpc3} the  trajectories of the projections of
the unit spin vector $\be_s$ and the unit vector in the direction of
the Cassini state $\be_c$ on the orbit plane over the $3\times 10^6$ year
interval covered by the Quinn data. Initially, the two unit vectors
are coincident ($\delta(t=0)=\delta_0=\cos^{-1}(\be_s\cdot \be_c)=0$).
That the two trajectories are almost indistinguishable
throughout the time interval shows how well the spin axis follows the
Cassini state for variations in the parameters that
define the state.  More precisely, we show in
Fig. \ref{fig:quinnadiabat4} the actual variations in $\delta(t)$ for
initial angular separations of $0^{\pr\pr}$ and $10^{\pr\pr}$. 
The results for
panel $a$ were obtained with spline fits through all 6185 data points
from the Quinn simulation, from which
$e(t),\,I(t),\,\Omega(t)\,dI(t)/dt$ and $d\Omega(t)/dt$ were
determined at each call to the Bulirsch-Stoer integrator and at points
therein. For panel $b$ a spline fit through only every other data
point was used including the first and last point of the full data
set. For the same initial conditions, the deviations of the spin axis
from the  Cassini state in panel $a$ are more scattered than those in
panel $b$ and reach a maximum of $2.8^{\pr\pr}$, whereas the maximum
deviation in panel $b$ is only $1.2^{\pr\pr}$, which difference is
explained as follows.

The interval between data points in the Quinn simulation is 180,000
Julian days or approximately 493 years. The shortest period covered in
a Fast Fourier Transform (FFT) of the data is the Nyquist period of
986 years (e.g. Press et al. 1986), which is less 
than the 1000-1100 year spin precession period. Although  the
spectral power at periods less than 2000 years are suppressed by as
much as 9 orders of magnitude in the Quinn data (Quinn et al. 1991),
we show in Section \ref{sec:standish} that the system is extremely
sensitive to variations in the inclination at periods near the spin precession
period, which period is included in the full data set. An FFT of
$dI/dt$ constructed from the spline fit to the full data set sampled
every 200 years, with either Hanning or Welch window functions, 
yields some power down to the 984 year Nyquist period, which includes
the precession period.

If we select only every other data point from the simulation, the  
Nyquist period is increased to 1972 years, which now excludes the
resonant period.  The FFTs of the spline fit and its derivative
through the inclination 
points of the half data set sampled at 200 years reproduces the
spectral power distribution of the full data set, but drops by more
than two orders of magnitude for periods shorter than the 1972 year
Nyquist period of the half data set.  In other words, the half data set
contains more than 100 times less power at the resonant period than
the full data set, and this accounts for the more erratic behavior and
larger maximum separation in panel $a$.  We show in Section
\ref{sec:standish} that there is little or no spectral power near a
period of 1000 to 1100 years in the full Mercury ephemeris. So the
separations of the spin from 
the Cassini state indicated in panel $b$, where the spin is initially
in the Cassini state, is more representative of the true limits on
this separation as the orbital elements vary according to the Quinn
solar system simulation over the 3 million year interval. Panel $c$
shows that the spin remains within about $1^{\pr\pr}$ of an initial
separation of $10^{\pr\pr}$ indicating that the adiabatic invariant is
reasonably well preserved.

The source of the maximum separation of the spin from the Cassini state in
panel $b$ of Fig. \ref{fig:quinnadiabat4} can be inferred by holding
$d\Omega/dt$ constant with precession period of 300,000 years and varying 
$I$ and $e$ according to  
\begin{eqnarray}
I&=&I_0+A_{\s I}\sin{\left(\frac{2\pi t}{P_{\s I}}\right)}, \nonumber\\ 
e&=&e_0+A_e\sin{\left(\frac{2\pi t}{P_e}+\phi\right)},\label{eq:dIedt}
\end{eqnarray}
and use the Quinn data to assign mean values $I_0=6.0^\circ,\,e_0=
0.19$ and the following amplitudes with associated periods:
\begin{eqnarray}
A_{\s I}=5.5^\circ&{\rm for}&P_{\s I}=1\times 10^6\,{\rm yr},\nonumber\\
A_{\s I}=1.5^\circ&{\rm for}&P_{\s I}=2\times 10^5\,{\rm yr},\nonumber\\
A_{\s I}=0.6^\circ&{\rm for}&P_{\s I}=5\times 10^4\,{\rm yr},\nonumber\\ 
A_e=0.06&{\rm for}&P_e=8\times 10^5\,{\rm yr},\nonumber\\
A_e=0.01&{\rm for}&P_e=5\times 10^4\,{\rm yr}. \label{eq:amperiod}
\end{eqnarray}
\begin{figure}[ht]
\epsscale{.6}
\plotone{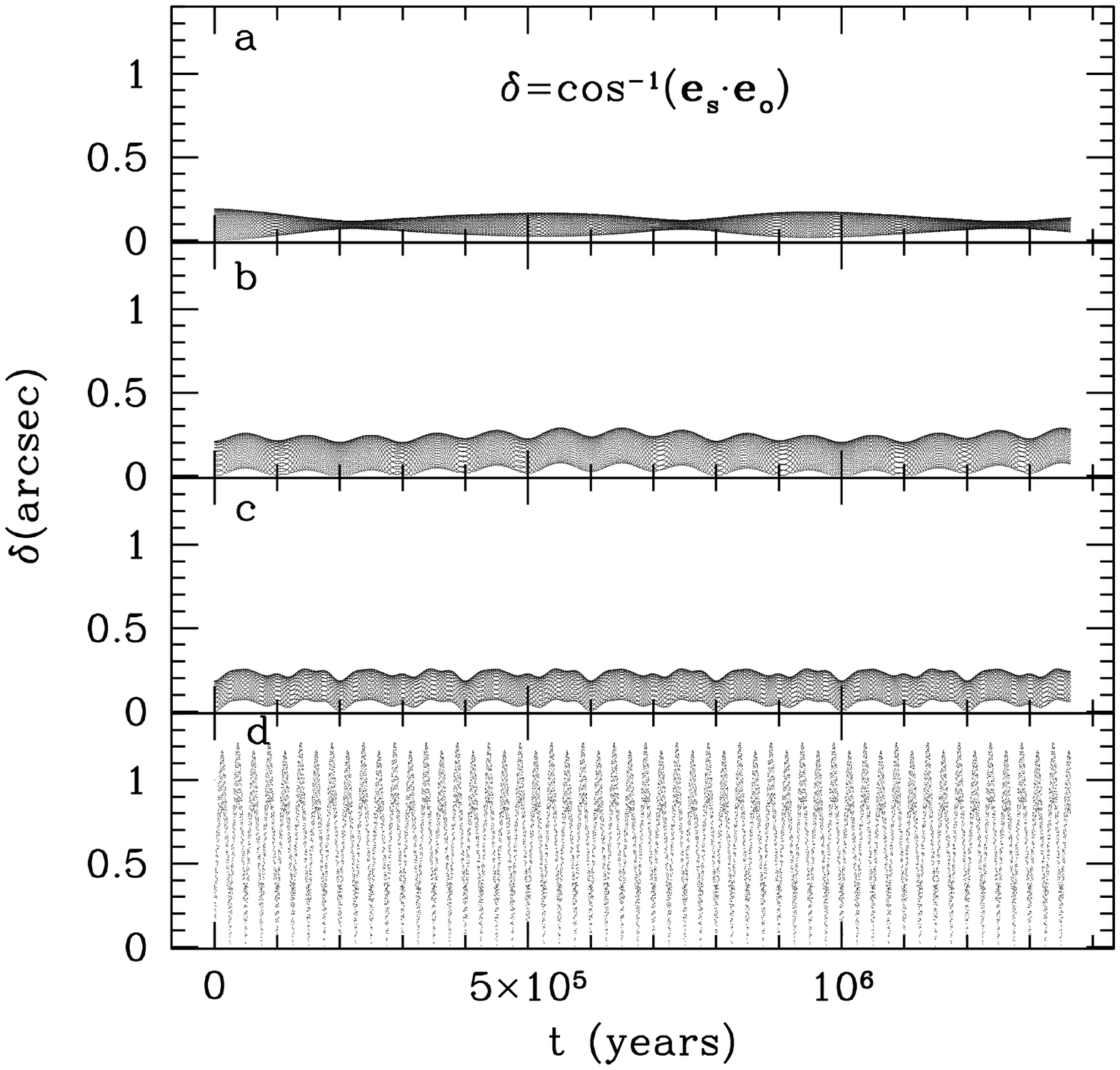}
\caption{Separation of spin and Cassini state for periodic variations
$I=I_0+A_{\s I}\sin{(2\pi t/P_{\s I})}$,
$e=e_0+A_e\sin{(2\pi t/P_e+\phi)}$, where initially the spin axis and
Cassini state are coincident. The amplitudes and periods are
characteristic of those in the Quinn data. $I_0=6.0^\circ$ and
$e_0=0.19$ are the real approximate mean values. These mean values and
$\phi=0$ are assumed for all
the cases. a) $A_{\s I}=5.5^\circ,\,P_{\s I}= 10^6\,{\rm
yr},\;A_e=0.06,\,P_e=8\times 10^5\,{\rm yr}$;  
b) $A_{\s I}=1.5^\circ,\,P_{\s I}= 2.0\times 10^5\,{\rm
yr},\;A_e=0.06,\,P_e=8\times 10^5\,{\rm yr}$; c)  
$A_{\s I}=1.5^\circ,\,P_{\s I}=2\times 10^5\,{\rm yr},\;A_e=0.01,\,
P_e=5\times 10^4\,{\rm yr}$; d) $A_{\s I}=0.6^\circ,\,P_{\s I}=5\times
10^4\,{\rm yr},\;A_e=0.01,\,P_e=5\times 10^4\,{\rm yr}$.   
\label{fig:spincassperiodic}}
\end{figure}

\begin{figure}[ht]
\epsscale{.6}
\plotone{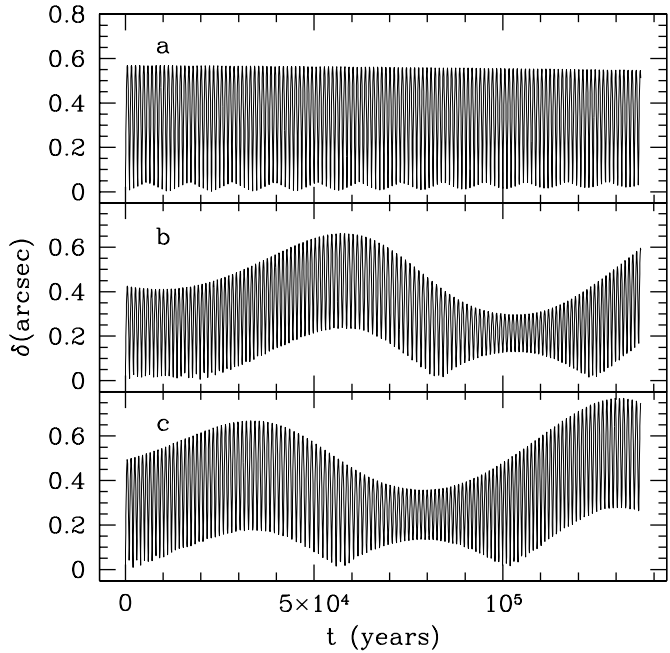}
\caption{Separation of spin and Cassini state for the maximum $dI/dt$
in the spline fit to the Quinn data. $I=7.35^\circ+ ({1.2\times
10^{-4}}^\circ/{\rm yr})t$. a) $e=0.19$. b) $e=0.19+0.06\sin{2\pi
t/10^5{\rm yr}}$. c) $e=0.19+0.06\sin{[2\pi t/10^5{\rm yr}+
90^\circ]}$ \label{fig:linearI}}  
\end{figure}

Fig. \ref{fig:spincassperiodic} shows the results for $\delta(t)$ for
$e$ and $I$ variations with several combinations of the above
amplitudes and periods. The initial value $\delta_0=0$ is chosen for
all cases. The large amplitude, long period variations lead to
maximum deviations of the spin from the Cassini state of about
$0.25^{\pr\pr}$, whereas shorter period variations ($P=5\times
10^4\,{\rm yr}$) with smaller amplitude lead to a maximum deviation a
little more than $1^{\pr\pr}$. It is these shorter period variations in
the the orbital parameters that are causing the maximum deviations of
the spin from the Cassini state in Fig. \ref{fig:quinnadiabat4}.
The fact that different values of $A_e$ with $P_e=5\times 10^4\,{\rm
yr}$ in combination with $A_{\s I}=1.5^\circ$ with $P_{\s I}=2\times
10^5\,{\rm yr}$ yield comparable maximum values of $\delta(t)$ shows
that variations in $I$ are much more important than the variations in
$e$ in causing $\be_s$ to not follow the Cassini state.  Next, the
maximum deviation from the Cassini state is almost proportional to
$A_{\s I}$ for a fixed $P_{\s I}$, from which we infer that it is
really the maximum $dI/dt$ that determines how much $\be_s$ deviates
from $\be_c$.  If we set
$I=I_0+(dI/dt)_{max}t$, with $(dI/dt)_{max}=1.2\times
10^{-4}\,^\circ/{\rm yr}$ from the spline fit to the Quinn data, the
excursions of the spin away from the Cassini state over approximately
$3.5\times 10^5$ years are shown in Fig. \ref{fig:linearI}. For $e\equiv
0.19$, the maximum $\delta<0.6^{\pr\pr}$. Imposing a periodic variation
in $e$ at the maximum amplitude of 0.06 while $I$ is increasing
linearly, induces a modulation in the maximum $\delta$, whose peaks
approach only to $0.8^{\pr\pr}$. 

\section {Results for short period variations \label{sec:standish}}
\begin{figure}[ht] 
\epsscale{.5}
\plotone{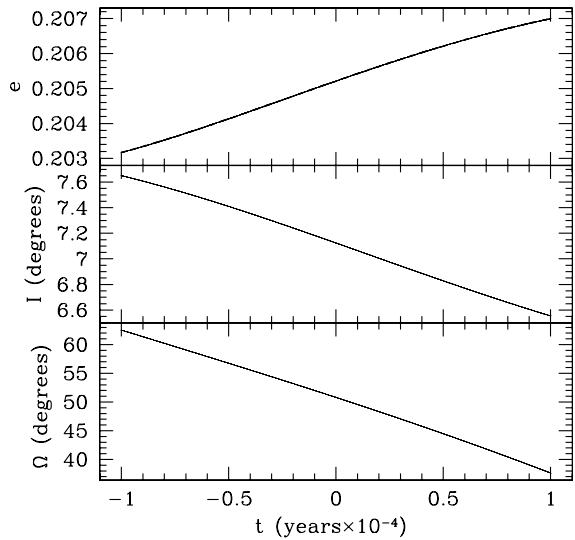}
\caption{Variation of $e,\,I$  and $\Omega$ relative to the ecliptic of J2000
for JPL Ephemeris DE 408. \label{fig:standish1}} 
\end{figure}

The Quinn data has been filtered to eliminate all variations with
periods less than 2000 years, so it is instructive to look at the
response of the system to short period fluctuations. For this purpose
we use the complete Mercury orbital element variations, sampled every 500
Julian days from the JPL Ephemeris DE 408, that were kindly provided
by Myles Standish. First we determine the location of the current
Cassini state. Fig. \ref{fig:standish1} shows the variation over 
20,000 years of $e,\,I$ and $\Omega$ relative to the ecliptic plane,
of J2000, where $\Omega$ is measured from the vernal equinox. The
variations are dominated by nearly linear secular changes with short
period fluctuations superposed whose amplitudes are within the line
widths of the curves.  
The current position of the Cassini state is necessary for the
interpretation of the radar and future spacecraft information.  
We can determine a preliminary position of the Cassini state that will
be refined when the MESSENGER spacecraft orbits Mercury. The
obliquity of the state is given by Eq. (\ref{eq:ic}). For the
value of $w$ chosen for the equations of motion determinations, the
precession period would be 286,660 years and $I^\pr=7.51^\circ$,
whereas for the real value of $w=w_{\s L}$ relative to the Laplace
plane normal, 
the precession period would be 328,000 years and $I^\pr=8.6^\circ$
(Yseboodt and Margot, 2005).  Both sets yield
$w\sin{I^\pr}=1.641\times 10^{-4}\,{^\circ}/{\rm year}$ in the numerator of
Eq. (\ref{eq:ic}). With $C/MR^2=0.34$, a central value of a plausible
range (Harder and Schubert, 2001), $e=0.206$, $n=2\pi/(87.969\,{\rm
d})$, 365.2563 d/y, $(J_2,\,C_{22})=(6.0\times
10^{-5},\,1.0\times 10^{-5})$ (Anderson {\it et al.} 1987),
$wR^\pr=0.1408^\circ/{\rm year}$ and $wS=0.01437^\circ/{\rm
year}$ (Recall $R^\pr$ and $S$ have $w$ in their denominators.), the two
choices of $w$ and corresponding $I^\pr$ yield  
$i_c=1.6704^\pr\,{\rm and}\, 1.6696^\pr$ respectively, a difference of
only $0.05^{\pr\pr}$, which justifies the use of $w$ in place of
$w_{\s L}$ in the determinations of the Cassini state.  Yseboodt and
Margot (2005) find $i_c=1.68^\pr$. If we use the
published uncertainties in $J_2$ and $C_{22}$ of $\pm 2\times 10^{-5}$ and
$\pm 0.5\times 10^{-5}$ respectively and assume that the uncertainties
in the ephemerides used to determine $I^\pr$ and the uncertainty in
$C/MR^2$ are negligible, $1.18<i_c<2.51$ arcmin.  

The unit vector
corresponding to the Cassini state position $\be_c$ is in the
plane determined by the orbit normal $\be_o$ and the $\be_{\s L}$ on
the opposite side of $\be_o$ from $\be_{\s L}$.  From the constraints
that $\be_c\cdot (\be_o\times\be_{\s L})=0$ and
$\be_c\cdot\be_o=\cos{i_c}$ and the unit magnitude of $\be_c$, three
equations in the three components of $\be_c$ can be solved for the current
position of the Cassini state.  It is less algebraically taxing to
solve for the increments $\Delta X$ and $\Delta Y$ relative the
projection of $\be_o$ on the ecliptic plane to yield the ecliptic
latitude and longitude of the Cassini state position of
$\lambda_c=82.9694^\circ$ and $\phi_c=-41.7585^\circ$ respectively for
the J2000 epoch with $i_c=1.67^\pr$. The direction of the displacement
is found from the 
intersection of the plane determined by $\be_o$ and any of the $\vec w$
from Eq. (\ref{eq:vecw1}) and the unit sphere. This position will of
course have to be adjusted as the uncertainties in the determining
parameters are reduced. 

\begin{figure}[ht]
\epsscale{.6}
\plotone{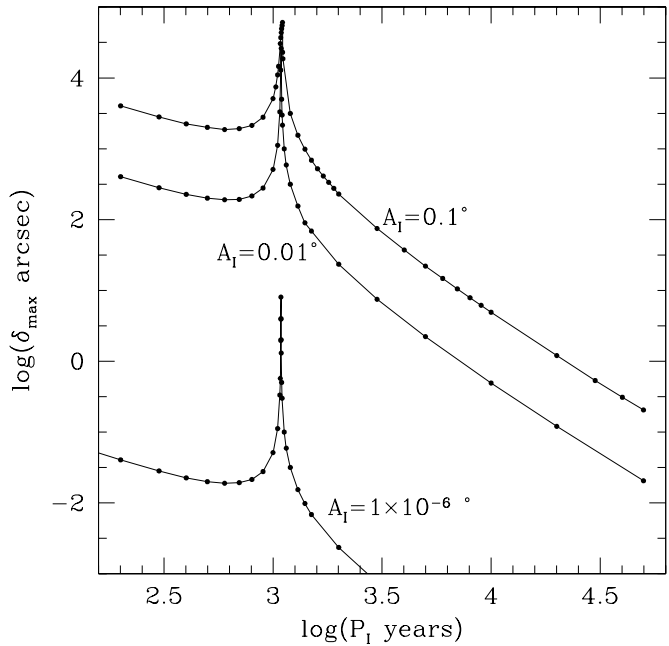}
\caption{Maximum separation of the spin axis from the Cassini state
for periodic variation of $I$ with amplitude $A_I$ with $e=0.19$ and
$d\Omega/dt=-2\pi/(287,000\,{\rm years})$ as a
function of the period of variation. \label{fig:delmaxvsiperiod}}
\end{figure}

The response of the angular separation $\delta$ of spin axis $\be_s$
and the Cassini state $\be_c$ to short period fluctuations can be
investigated by fixing $e=0.19$ and $d\Omega/dt=-2\pi/(287,000\,{\rm
years})$ but forcing $I$ to vary as  $I=I_0+A_{\s I}
\sin{(2\pi t/P_{\s I})}$. We fix the amplitude $A_{\s I}$, and determine
the maximum separation of the spin from the Cassini state as a
function of the period of the variation  $P_{\s I}$.  Panel $d$ in Fig.
\ref{fig:spincassperiodic} is characteristic of the behavior of the
spin-Cassini state separation for periodic variations in $e$ and
$I$. The separation starting at $\delta_0=0$ fluctuates between zero
and a maximum, which is about $1.2^{\pr\pr}$ in panel $d$ in Fig. 
\ref{fig:spincassperiodic}.  Fig. \ref{fig:delmaxvsiperiod} shows this
maximum separation for periods ranging from 200 years to $5\times
10^4$ years for several values of $A_{\s I}$.  Quite large separations
of the spin from the Cassini state can result from such relatively
short period oscillations in $I$, but most noticeably at periods near
the period of the spin precession about the Cassini state, where a clear
resonant response is evident. 

We saw in Section \ref{sec:quinn} how a 
very small spectral power in the Quinn data at a period near the
resonance more than doubled the maximum spin-Cassini state separation
over the 3 million year interval from the $1^{\pr\pr}$ maximum
separation when spectral power at that period was suppressed by
halving the number of data points.  To check whether there is
significant power in the real variations in $e,\,I,\,\Omega,\,dI/dt$ and
$d\Omega/dt$ at the resonant period or at any other short period that
could cause the spin 
to separate from the Cassini state by a large angle, we repeat the
exercise in Section \ref{sec:quinn}, but now with the variations in
$e,\,I\,{\rm and}\,\Omega$ and their derivatives given by the
ephemeris DE 408. 
Fig. \ref{fig:standish1} shows that the short period fluctuations in
$e$, $I$ and $\Omega$ are small compared to  significant almost
linear variations.

As was done with
the Quinn data, we represent the variations in $e$, $I$, and $\Omega$
with spline fits that capture the short period fluctuations in these
elements as well as their derivatives derived therein for use in
Eqs. (\ref{eq:dpqdt}).  The Cassini state position is determined at
arbitrary times from averaged values of the parameters, with $\langle e
\rangle,\, \langle I\rangle,\,{\rm and}\, \langle\Omega\rangle$ being
averages determined by the sum of the values at the extremes of a 2000
year window centered on the epoch divided by 2, and 
$\langle dI/dt\rangle$ and $\langle d\Omega/dt\rangle$ 
are the difference in the values at the extremes divided by 2000 years.
\begin{figure}[ht]
\epsscale{.6}
\plotone{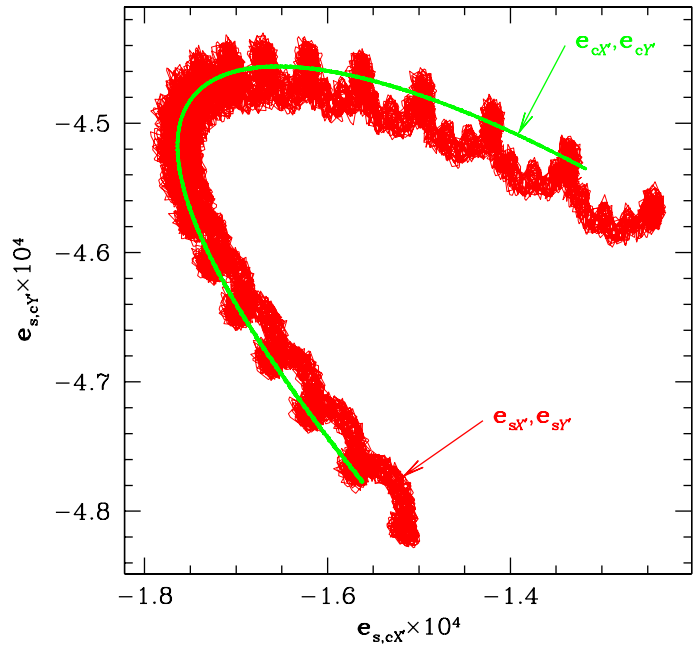}
\caption{Trajectories of the projections of the unit spin vector
$\be_s$ and the unit vector in the Cassini state direction $\be_c$ on
the orbit $X^\pr Y^\pr$ plane for variations in the orbital elements
according to the 20000 year JPL Ephemeris DE 408.  \label{fig:qp_qcpc4}}
\end{figure}

\begin{figure}[ht]
\epsscale{.6}
\plotone{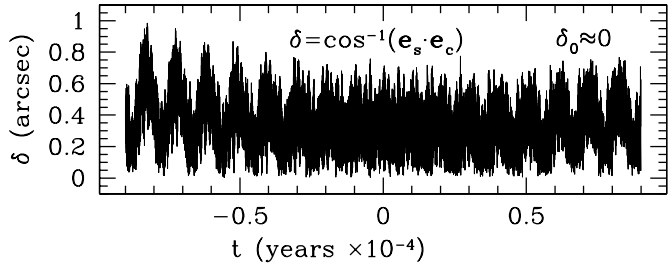}
\caption{Separation $\delta$ of the spin vector from 
Cassini state 1 for variations in $e,\,I$ and $\Omega$ from 
JPL ephemeris DE 408 covering 20,000 years centered on
calendar year 0 and sampled every 500 ephemeris days. 
The spin is initially at Cassini state 1 in
this example, and the separation is deduced from a spline fit through
the 14,624 data points of the ephemeris. The periodicity in $\delta$
is that of the free precession of the spin about the Cassini
state. \label{fig:de408del}} 
\end{figure}

In Fig. \ref{fig:qp_qcpc4}, the equivalent of Fig. \ref{fig:qp_qcpc3}
the projections of the components of $\be_s$
and $\be_c$ onto the orbit plane are shown over the 20,000 year
interval of the ephemeris, where the projection of $e_c$ is truncated
at the endpoints because of limits of the averaging process.
Fig. \ref{fig:qp_qcpc4} represents a short segment of one of the loops
in Fig. \ref{fig:qp_qcpc3} extended into the future to J10000, except
now all of the high frequency terms are included in the variations of
$e,\,I$ and $\Omega$ for the variation of the spin position. Again the
spin is initially in the Cassini state 
($\delta_0=0$), and we see that it remains close to that state as we
have defined it in terms of the averages of the ephemeris
data. In Fig. \ref{fig:de408del}, the equivalent of
Fig. \ref{fig:quinnadiabat4}b, we show that the spin, initially in the
Cassini state, stays within
$1^{\pr\pr}$ of the Cassini state over the 20,000 year interval of the
ephemeris. There are no 
short period terms in the real variations of $e$ and $I$ that can lead
to the large periodic separations shown in
Fig. \ref{fig:delmaxvsiperiod}. In particular, there are no
contributions to the variations that are near resonance with the spin
precession, and the spin will remain close to the Cassini state for
both long period and short period variations in the orbital elements
and variations in the Laplace plane orientation that define the
position of the state.  
   
\section{Discussion \label{sec:discussion}}
We have shown that Mercury's spin axis will stay within approximately
$1^{\pr\pr}$ of the time varying position of Cassini state 1 after
dissipative processes have brought it to the state for both long and
short period variations in the orbital parameters. Yseboodt and Margot
(2005) also find that Mercury's spin axis remains close to the Cassini
state as the position of the state is altered by slowly increasing the
planetary masses from zero to their current values with time scales
that are long compared to the spin precession period.  The maximum
separation of $1^{\pr\pr}$ that we found is about 1\% of the
$1.67^{\pr}$ current obliquity of the Cassini state. Current
estimates for the precision of the MESSENGER spacecraft determination
of Mercury's  
obliquity are about 10\% of its value (Zuber and Smith, 1997),
although this precision may possibly be  exceeded by radar
measurements.  In any case the maximum deviation of Mercury's
obliquity from the Cassini state obliquity induced by the slow and
fast orbital variations and by the changing geometry of the solar
system is sufficiently small that reasonably precise values of
$C/MR^2$ will be obtainable from the radar and spacecraft observations
with correspondingly tight constraints on $C_m/C$. This conclusion
depends of course on there being no recent excitation of a free
spin precession. With tight constraints on both $C/MR^2$ and $C_m/C$, the
internal structure of Mercury should be reasonably well constrained.

For the small coupling that is likely between Mercury's liquid
core and solid mantle, the core is not likely to follow the mantle on
the spin precession time scale as we have assumed here. If the mantle
precesses independently of the core, the polar moment of inertia for
the spin precession will be about half of the total moment of inertia. The
spin precession period will be 500 years instead of the 1000 years assumed
here. The mantle would relax to a somewhat different Cassini state
which would gradually relax to the current state as dissipative
processes cause the liquid core to catch up on a time scale short
compared to the 300,000 year orbit precession time scale. 
It may be the case that there is a slight offset of the spin axis from
the Cassini state because of the fluid coupling between core and
mantle.  If this offset is measurable, it will provide a constraint
on the core-mantle coupling.  Investigation of this conjecture will
be included in a following paper, where core and mantle are considered
as independent but coupled entities.
\section{Acknowledgements}
It is a pleasure to thank Man Hoi Lee and Gerald Ramian for very
helpful discussions and M. H. Lee for some informative numerical
calculations. Critical reviews by Jean-Luc Margot and Jacques Henrard
motivated changes in presentation that greatly improved the
manuscript. Marie Yseboodt and Jean-Luc Margot pointed out a serious
error introduced into the revised version of the manuscript. Thanks
are also due Tom Quinn for providing the files 
from his 3 million year simulation of the solar system and Myles
Standish for kindly providing Mercury's orbital variations in a
convenient format from his just completed 20,000 year JPL DE 408 ephemeris.
This work is supported in part by the Planetary Geology and Geophysics
Program of NASA under grant NAG5 11666 and by the MESSENGER mission
to Mercury.

\section{References:}
\parindent=0pt
Anderson, J.D., Colombo, G., Esposito, P.B., Lau, E.L., Trager, T.B.
1987. The mass, gravity field and ephemeris of Mercury. Icarus
71, 337-349. 

Anselmi, A., Scoon, G.E.N. 2001. BepiColombo, ESA's Mercury
Cornerstone mission. Planet. Space Sci. 49, 1409-1420.

Beletskii, V.V. (1972) Resonance rotation of celestial bodies and
Cassini's laws. Cel. Mech. 6, 356-378. 

Colombo, G. (1966) Cassini's second and third laws, {\it Astron. J.}
{\bf 71}, 891-96.

Goldreich, P., Toomre, A. 1969. Some remarks on polar wandering,
J. Geophys. Res. 74, 2555-2567.

Harder, H., Schubert, G.  2001. Sulfur in Mercury's core? 
Icarus  151, 118-122.

Kaula, W.M. 1966  Theory of satellite geodesy; applications of
satellites to geodesy, Blaisdell, Waltham, MA.

Margot, J.L., Peale, S.J., Slade, M.A., Jurgens, R.F., Holin, I.V., 
2003. Mercury interior properties from measurements of librations. 
Mercury, 25th meeting of the IAU, Joint Discussion 2, 16 July, 2003, 
Sydney, Australia, meeting abstract. (http://adsabs.harvard.edu).

Peale, S.J. 1969. Generalized Cassini's laws, Astron. J. 74, 483-89.

Peale, S.J. 1974. Possible histories of the obliquity of Mercury,
Astron. J. 79, 722-44.
 
Peale, S.J. 1976. Does Mercury have a Molten Core? 
Nature 262, 765-766.

Peale, S.J. (1981) Measurement accuracies required for the
determination of a Mercurian liquid core, {\it Icarus} {\bf 48},
143-45.

Peale, S.J.  1988. Rotational dynamics of Mercury and the state of its
core, In  Mercury, ed. F. Vilas, C.R. Chapman, M.S. Matthews,
U. of Arizona Press, Tucson, 461-493.

Peale, S.J. 2005. The free precession and libration of Mercury,
Icarus, In press.  

Peale, S.J. Phillips, R.J., Solomon, S.C., Smith, D.E., Zuber, M.T.,
 2002. A procedure for determining the nature of Mercury's core, 
Meteor. Planet. Sci.  37, 1269-1283.

Press, W.H., Flannery, B.P., Teukolsky, S.A., Vetterling, W.T. {\it
Numerical Recipes} Cambridge University Press, London, p. 86.

Quinn, T.R., Tremaine, S., Duncan, M. 1991. A three million year
integration of the Earth's orbit, Astron. J. 101, 2287-2305.

Solomon, S.C., 20 colleagues  2001. The MESSENGER mission to Mercury:
Scientific objectives and implementation,  Planet. Space Sci.
 49, 1445-1465.

Ward, W.M. 1975. Tidal friction and generalized Cassini's laws in the
solar system,  Astron. J. 80, 64-70.

Yseboodt, M. and Margot, J.L. 2005. Evolution of Mercury's Obliquity,
Submitted to Icarus.

Zuber M. T. and Smith D. E. 1997. Remote sensing of
planetary librations from gravity and topography data: Mercury
simulation (abstract). {\it Lun. Planet. Sci.} {\bf 28}, 1637--1638.

\end{document}